\UseRawInputEncoding
\documentclass[journal=jpcl,manuscript=article, layout=twocolumn]{achemso}
\usepackage[version=3]{mhchem} % Formula subscripts using \ce{}
\usepackage{xcolor}
\usepackage[capitalise]{cleveref}
\crefname{equation}{reaction}{reactions}
\Crefname{equation}{Reaction}{Reactions}
\usepackage{multirow}
\usepackage[switch, left]{lineno}
\usepackage[colorinlistoftodos,prependcaption,textsize=small]{todonotes}
\author{Yuebei Xiong}
\affiliation{Institute of Quantum Precision Measurement, College of Physics and
Optoelectronic Engineering, Shenzhen University, Shenzhen 518060,
China}
\author{Zhirui Gong}
\affiliation{Institute of Quantum Precision Measurement, College of Physics and
Optoelectronic Engineering, Shenzhen University, Shenzhen 518060,
China}
\email{gongzr@szu.edu.cn}
\author{Hao Jin}
\affiliation{College of Physics and Optoelectronic Engineering, Shenzhen University,
Shenzhen 518060, P. R. China}
\email{jh@szu.edu.cn}

\title{Strain tuning of the nonlinear anomalous Hall effect in $\textrm{MoS}_{2}$ monolayer}
\abbreviations{IR,NMR,UV}

\begin{document}

\begin{abstract}
Due to the time reversal symmetry, the linear anomalous Hall effect
(AHE) usually vanishes in $\textrm{MoS}_{2}$ monolayer. In contrast, the nonlinear AHE
plays an essential role in such system when the uniaxial strain breaks the $C_{3v}$ symmetry and
eventually results in the nonzero Berry curvature dipole (BCD). We find that not only the magnitude of the AHE
but also the nonlinear Hall angle can be tuned by the strain. Especially 
the nonlinear Hall angle exhibits a deep relationship which is analogy to
the birefraction phenomenon in optics. It actually results from the
pseudotensor nature of the BCD moment. Besides
the ordinary positive and negative crystals in optics, there are two
more birefraction-like cases corresponding to an imaginary refraction
index ratio in monolayer $\textrm{MoS}_{2}$. Our
findings shed lights on the strain controlled electronic
devices based on the two-dimensional (2D) materials with BCD.
\end{abstract}

The Berry phase of the electronic wave function plays a profound role
on material properties \cite{1,2}. The adoption of the Berry curvature (BC)
concept establishes the link between the topological nature of electron
motion and the intriguing phenomena such as spin/valley Hall effects \cite{3,4,5,6},
optical selection rules \cite{7,8} and magnetoelectric effects \cite{9,10,11}.
Among these phenomena, the various Hall effects are exclusively studied,
which facilitates the manipulation of the electronic currents with
different degree of freedoms such as charge, spin and valley indices \cite{12,13}.
In two-dimensional (2D) materials without external magnetic field, only the out-of-plane
part of the BC survives, which leads to a simple
form of the anomalous Hall effect (AHE) \cite{14}. Since the first
order Hall conductivity vanishes in time-reversal symmetric crystals,
the second order one should be taken into consideration profoundly
relating to the Berry curvature dipole (BCD) \cite{15}, whose definition
is the first-order moment of the BC over the occupied
electronic states \cite{16}. It has been studied in various systems
such as 2D system \cite{17,18,19,20,21,22,23,24,25},
the Moire superlattice \cite{26} and the disorder system \cite{27}.
Different nonlinear currents such as the nonlinear Hall currents and
the optical Hall currents are investigated based on the relaxation
time approximation \cite{28}. For the highly symmetric system, higher
order of the nonlinear AHE are also taken into consideration \cite{29,30}.
In contrast to the scalar linear Hall conductivity requiring the time-reversal
broken, the nonlinear Hall conductivity is a pseudotensor and can
survive in the solid even possessing the time-reversal symmetry \cite{31,32,33}.

Recently, the 2D transition metal dichalcogenides (TMDs) monolayer such as
$\textrm{MoS}_{2}$ attract a lot interest according to their unique
electronic and optical properties in the 2D limit and promising application
prospects \cite{34,35,36}. As a naturally spatial inversion symmetry broken
system, there are quite a lot of intriguing phenomena relating to the nonzero
 BC \cite{37}. The nonlinear AHE also vanishes
in such system because under the $C_{3v}$ symmetry the BCD is exactly zero. In
this sense, the uniaxial strain plays an essential role to achieve considerable nonlinear AHE.
Moreover, in contrast with the linear Hall
angle which is fixed at 90 degrees, the nonlinear Hall angle can be tuned
by the strength of the strain and the orientation of the strain
axis \cite{38,39,40}. However, the hidden relationship between the nonlinear
Hall angles, the orientation of uniaxial strain and the orientation of the ac electric field has not been revealed yet.

Here, we study the uniaxial strain tuned nonlinear AHE
in the $\textrm{MoS}_{2}$ monolayer. By applying ac electric field and
the uniaxial strain, the nonlinear AHE can be achieved in such system.
We find that the nonlinear Hall angle, the orientation of the strain axis and
the orientation of the ac electric field exhibit
a deep relationship which is analogy to the birefraction phenomenon
in optics. The orientation of the ac electric field and the nonlinear Hall angle plays the roles as the ordinary
and the extraordinary lights in the optical counterparts. Just like
the birefraction resulting from the dielectric constant tensor, the
hidden relationship between the nonlinear Hall angles, the orientation of the strain and the ac electric field can
be established due to the pseudotensor nature of the BCD. In this sense, there are effective refraction indices for
these birefraction-like phenomenon in $\textrm{MoS}_{2}$.
Besides the ordinary positive and negative crystals in optics, we
also find that there are two more birefraction-like cases corresponding
to imaginary refraction indices in monolayer $\textrm{MoS}_{2}$
when accurately tuning the uniaxial strain. Our findings can be generalized
to other 2D systems and shed lights on the strain tuning
electronic devices based on the 2D materials.

%%%%%%%%%%%%%%%%%%%%%%%%%%%%%%%%%%%%%%%%%%%%%%%%%%%%%%%%%%%%%
\begin{figure}[h]
\centering
\includegraphics[width=8cm]{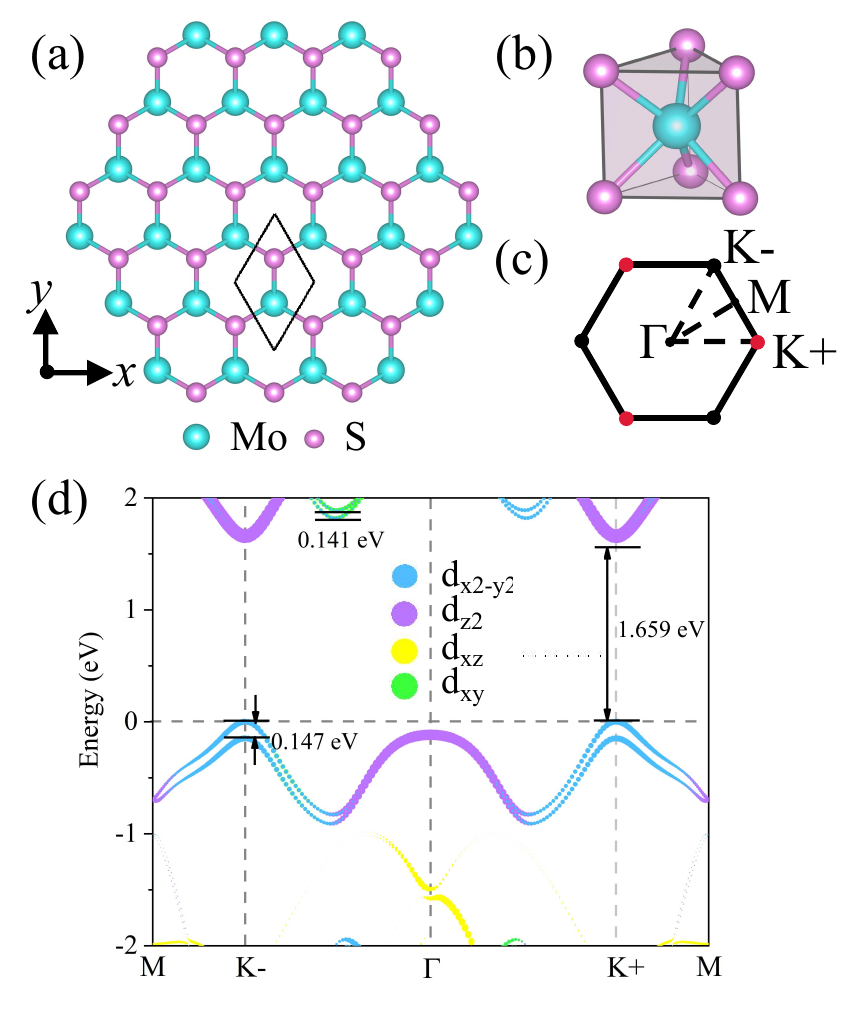}
\caption{(a) The top view of $\textrm{MoS}_{2}$  monolayer. The diamond region shows
the 2D primitive cell. (b) The schematic for the structure
of trigonal prismatic coordination. (c) The 2D first Brillouin zone
with special k points. The two inequivalent valleys $\textrm{K}+$ and $\textrm{K}-$
and their equivalent counterparts are shown in red and black respectively. (d) The band structure of monolayer $\textrm{MoS}_{2}$ with SOC. Contributions from Mo d orbitals: blue dots for $d_{x^{2}-y^{2}}$, purple dots for $d_z^{2}$, yellow dots for $d_{xz}$, green dots for $d_{xy}$.}
\label{fig1}
\end{figure}

%%%%%%%%%%%%%%%%%%%%%%%%%%%%%%%%%%%%%%%%%%%%%%%%%%%%%%%%%%%%%
We focus on the origin of nonlinear AHE in the 2D TMDs monolayer, e.g., $\textrm{MoS}_{2}$  monolayer. The
crystal structure of $\textrm{MoS}_{2}$ monolayer is depicted in Fig.~\ref{fig1}(a). They are in trigonal prismatic coordination including a Mo atom and two S atoms, which is depicted in Fig.~\ref{fig1}(b). The primitive cell of ${\textstyle \textrm{MoS}_{2}}$ monolayer
shows the breaking of spatial inversion symmetry apparently. The direct band gaps of $\textrm{MoS}_{2}$ monolayer are located at $\textrm{K}+$ and $\textrm{K}-$ valleys. The Bloch states of $\textrm{MoS}_{2}$ monolayer near the band edges mostly consist of Mo $4d$ orbitals, especially the $d_{xy}$ , $d_{x^{2}-y^{2}}$, and $d_{z^{2}}$ orbitals, which are dominant components for conduction bands (CB) and valence bands (VB). The effective Hamiltonian
up to the nearest-neighbor hopping \cite{41}

\begin{equation}
H=I_{2}\otimes H_{0}+H_{soc}+H_{strain}\label{eq:1}
\end{equation}
contains three parts: the k.p Hamiltonian of d-orbital electron $H_{0}$,
the SOC term $H_{soc}$ and the strain induced term $H_{strain}$. The first term

\begin{equation}
H_{0}=\left[\begin{array}{cc}
\frac{\Delta}{2}+\varepsilon & at\left(q_{x}\tau-iq_{y}\right)\\
at\left(q_{x}\tau+iq_{y}\right) & -\frac{\Delta}{2}+\varepsilon
\end{array}\right]\label{eq:2}
\end{equation}
describes a massive Dirac electron, where $\Delta$ is the band gap,
$\varepsilon$ is the correction energy bound up with the Fermi energy,
$a$ is the lattice constant, $t$ is the hopping constant and $\overrightarrow{q}=\left(q_{x},q_{y}\right)$
is the momentum vector. Here, the CB and the VB are respectively dominantly composed of ${d_{z^2}}$ orbital and ${d_{-2}}$ orbital, where $d_{-2}=\frac{1}{\sqrt{2}}(d_{x^{2}-y^{2}}-id_{xy})$. The valley dependent SOC term is written as
\begin{equation}
H_{soc}=\tau\sigma_{z}\otimes\left[\begin{array}{cc}
\lambda_{c} & 0\\
0 & \lambda_{v}
\end{array}\right],\label{eq:3}
\end{equation}
where $\tau$ is the valley index, $\sigma_{z}$ is the z component
Pauli matrix, $\lambda_{c}\left(\lambda_{v}\right)$ represents the
effective SOC splitting in the band edge of CB and
 VB. The band structure of $\textrm{MoS}_{2}$ monolayer 
is depicted in Fig.~\ref{fig1}(d). The energy splitting between the majority spin of the valence band maximum (VBM) and the minority spin of the VBM is up to $2 \lambda_{v}=147$ meV. In contrast, The energy splitting between the majority spin and the minority spin of the conduction band minimum (CBM) is $2 \lambda_{c} \sim 0$ meV.  

Usually, BCD is non-zero in a crystal if the inversion symmetry is broken. However, the $C_{3v}$ symmetry of the crystal forces the BCD to vanish. The strain-induced term $H_{strain}=\tau at\overrightarrow{\alpha}\cdot\overrightarrow{q}$
is introduced to break the $C_{3v}$ symmetry, where $\overrightarrow{\alpha}=\left(u_{xx}-u_{yy},-2u_{xy}\right)$
containing both the uniaxial and the shear strain and $u_{ij}=\left(\partial_{i}u_{j}+\partial_{j}u_{i}\right)/2$
are the in-plane deformations. It can arise when applying uniaxial strain in the zigzag direction.\cite{42}
Here, we have ignored the strain induced
isotropic and anisotropic position-dependent Fermi velocity in comparison
with free Fermi velocity in $H_{0}$ \cite{40}.
%%%%%%%%%%%%%%%%%%%%%%%%%%%%%%%%%%%%
%%%%%%%%%%%%%%%%%%%%%%%%%%%%%%%%%%%%%%%%%%%%%%%%%%%%%%%%%%%%%
\begin{figure*}[h]
\includegraphics[width=\textwidth]{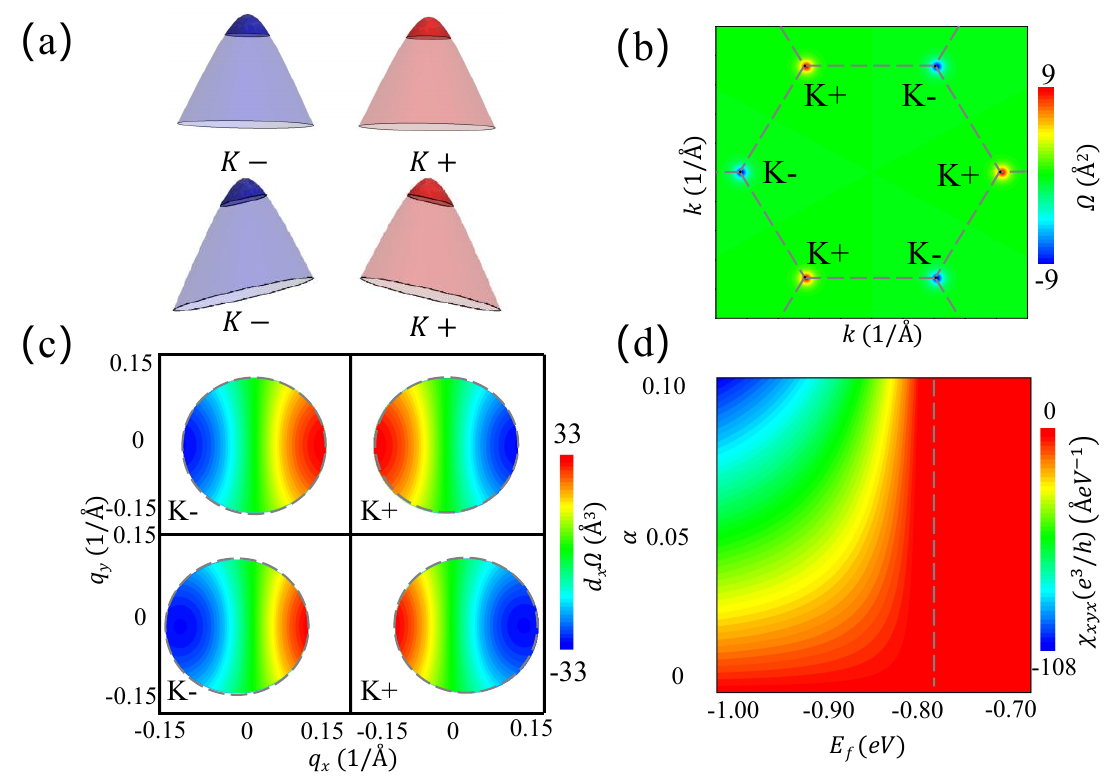}
\centering
\caption{(a) The Fermi pockets without strain and with strain. The darker blue and red parts denote the carrier distribution respectively corresponding to positive and negative BC, which gives rise to the Hall current. In order to
obtain the nonzero BCD, the tilted Dirac cone is
necessary and can be completely controlled by the strain which can be controlled by the strain. (b) The BC in the first Brillouin zone. (c) The $\partial_{x}\Omega$ without and with strain in different valleys. The grey dash lines represent the Fermi surface. 
(d) $\chi_{xyx}$ verses ${E_{f}}$ and $\alpha$. The parameters are $\Delta$ = 1.766 eV, a = 3.160 \text{\AA}, ${E_{f}}$ = 0.81 eV, t = 1.137 eV, $\lambda_{v}$ = 0.73 meV. The grey dash line represents the Fermi surface. }
\label{fig2}  
\end{figure*}
%%%%%%%%%%%%%%%%%%%%%%%%%%%%%%%%%%%%%%%%%%%%%%%%%%%%%%%%%%%%%

The nonlinear currents need to be taken into consideration which exactly
result from the BCD. In the former references \cite{15,16,18,20} only the scalar part
of the BCD is considered which gives rise to nonlinear
current as $j_{a}^{n}=\Re\left[\chi_{acd}\left(e^{2i\omega t}+1\right)\right]E_{c}E_{d}.$
Here, the superscript $n$ stands for nonlinear, $a,c,d=x,y$ are
the in plane directions, $\mathbf{E}\left(t\right)=\left( E_{x} \mathbf{x}+E_{y} \mathbf{y}\right)e^{i\omega t}$
is the ac electric field with frequency $\omega$. Here, $\mathbf{x}$ and $\mathbf{y}$ are
respectively the unit vector along $x$ and $y$ directions, $E_{x}$ and $E_{y}$ are the corresponding
electric field strengths.

Based on the relaxation time approximation of the electron
distribution, the Langevin equation with the relaxation time $\tau_{0}$
can be written as
\begin{equation}
\frac{\partial f}{\partial t}=\frac{e}{\hbar}E_{a}\left(t\right)\partial_{a}f-\frac{1}{\tau_{0}}\left(f-f_{0}\right).\label{eq6}
\end{equation}
Here, $E_{a}\left(t\right)$ is the component of the ac electric field
along $a$ direction, where $a=(x,y)$, and $f_{0}$ is the Fermi
distribution without the ac electric field. The nonlinear conductivities are given as

\begin{align}
\chi_{acd} & =-\epsilon_{abc}\frac{e^{3}\tau}{2\left(1+i\omega\tau\right)}D_{d},\label{eq5}
\end{align}
with $D_{d}=\int\partial_{d}\Omega_{b}\left(\mathbf{k}\right)f_{0}\left(\mathbf{k}\right)\frac{d\mathbf{k}}{\left(2\pi\right)^{2}}$. Here, 
the BC is straightforwardly obtained as 
\begin{align}
\Omega_{b}\left(\mathbf{\mathbf{q}\pm K}\right)=\pm\frac{2a^{2}t^{2}(\Delta-\lambda_{v})}{\left[4a^{2}t^{2}q^{2}+(\Delta-\lambda_{v})^{2}\right]^{\frac{3}{2}}},\label{eq6}
\end{align}
 and 
\begin{align}
\partial_{d}\Omega_{b}\left(\mathbf{\mathbf{q}\pm K}\right)=\mp\frac{24a^{4}q_{d}t^{4}(\Delta-\lambda_{v})}{\left[4a^{2}t^{2}q^{2}+(\Delta-\lambda_{v})^{2}\right]^{\frac{5}{2}}}.\label{eq7}
\end{align}
The BC is depicted in Fig.~\ref{fig2}(b), where obviously the BC is opposite in the opposite valleys.  
Here, $b=z$ for the 2D materials and $\epsilon_{abc}$ is the Levi-Civita
symbol. In this sense, the only four surviving elements of the nonlinear conductivity tensor are
$\chi_{xyx}=-\chi_{yxx}$ and $\chi_{xyy}=-\chi_{yxy}.$ Apparently, $\chi_{xyx}$ and $\chi_{xyy}$ are real ones in the Drude limit $\omega\tau<<1$. Since $D_{x}$ related to $\partial_{x}\Omega$ which is anti-symmetric along $x$ direction, the $D_{x}$ is nonzero when the Dirac cone is tilted along $x$ direction as shown in Fig.~\ref{fig2}(a). In this sense, $D_{x}$ is independent of the strain along $y$ direction, and accordingly $D_{y}$ is independent of the strain along $x$ direction.

The nonlinear current vanishes when no strain is applied because $\partial_{d}\Omega_{b}\left(\mathbf{k}\right)$ 
is odd function as $\partial_{d}\Omega_{b}\left(\mathbf{\mathbf{q}\pm K}\right)=-\partial_{d}\Omega_{b}\left(-\mathbf{\mathbf{q}\pm K}\right)$
while Fermi distribution is even one as $f_{0}\left(\mathbf{\mathbf{q}\pm K}\right)=f_{0}\left(\mathbf{-\mathbf{q}\pm K}\right).$ The $\partial_{d}\Omega_{b}\left(\mathbf{k}\right)$ with and without applying the strain in different valleys are depicted in Fig.~\ref{fig2}(c) , where the Kramers degeneracy is quite obvious. When the strain is applied, the $\partial_{d}\Omega_{b}\left(\mathbf{k}\right)$ in the $\textrm{K}+$ and $\textrm{K}-$ shifted in opposite directions, resulting in a larger BCD, which leads to a respectively large nonlinear Hall current.
Therefore the strain plays an essential role in the appearance of
the nonlinear current. The nonlinear current vanishes when no strain is applied because BCD
is odd function as $\partial_{d}\Omega_{b}\left(\mathbf{\mathbf{q}\pm K}\right)=-\partial_{d}\Omega_{b}\left(-\mathbf{\mathbf{q}\pm K}\right)$
while Fermi distribution is even one as $f_{0}\left(\mathbf{\mathbf{q}\pm K}\right)=f_{0}\left(\mathbf{-\mathbf{q}\pm K}\right).$
The large magnitude of the nonlinear
current results from the long relaxation time $\tau_{0}\sim1 ps$ \cite{43}.The $\chi_{xyx}$ verses ${E_{f}}$ and $\alpha$ is shown in Fig.~\ref{fig2}(d), the $\chi_{xyx}$ decreases when increasing the strain strength $\alpha$ for a fixed Fermi surface $E_{f}<E_{VBM}$. The similar relationship appears in $\chi_{xyy}$, which dependents on the strain strength $\beta$ in $y$ direction. 

%%%%%%%%%%%%%%%%%%%%%%%%%%%%%%%%%%%%%%%%%%%%%%%%%%%%%%%%%%%%%
\begin{figure*}[h]
\includegraphics[width=\textwidth]{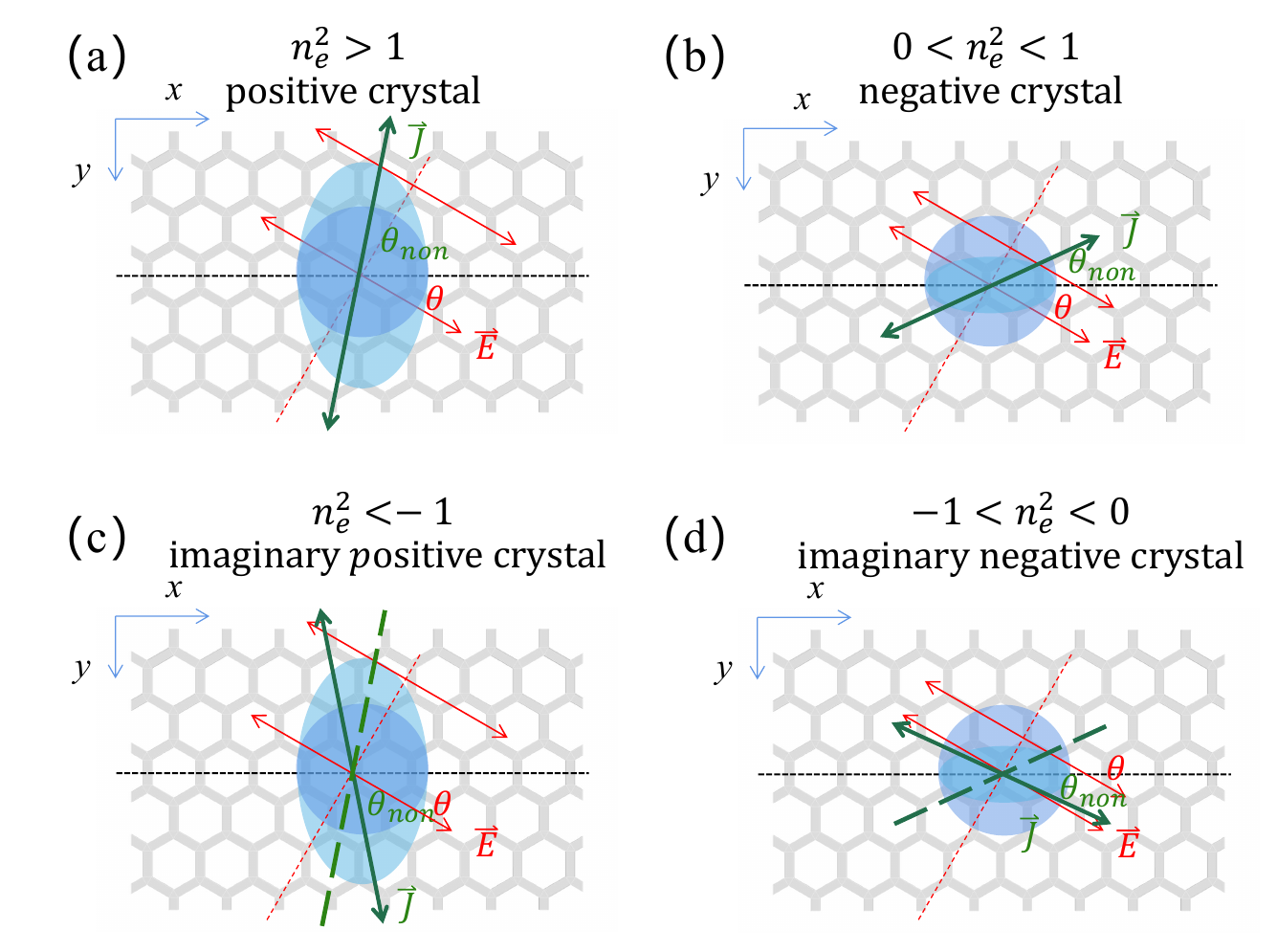}
\centering
\caption{Propagation direction of the nonlinear current determined by the Huygens' Principle for different $n_{e}^{2}.$ Here, The red arrow and darker green arrows represent the polarization direction of the electric field and the propagation direction of the nonlinear current. The darker green dash lines represents  the direction of the nonlinear current in (a) and (b) respectively.}
\label{fig3}
\end{figure*}

%%%%%%%%%%%%%%%%%%%%%%%%%%%%%%%%%%%%%%%%%%%%%%%%%%%%%%%%%%%%%
The corresponding
nonlinear current $\mathbf{j}=\left(j_{x},j_{y}\right)$
is given in the matrix form as
\begin{equation}
\left[\begin{array}{c}
j_{x}\\
j_{y}
\end{array}\right]=\Re\left[\left(e^{2i\omega t}+1\right)\overleftrightarrow{\chi}\left[\begin{array}{c}
E_{x}^{2}\\
2E_{x}E_{y}\\
E_{y}^{2},
\end{array}\right]\right]\label{eq7}
\end{equation}
the pseudotensor $\overleftrightarrow{\chi}=\left[\begin{array}{ccc}
0 & \chi_{xyx} & \chi_{xyy}\\
-\chi_{xyx} & -\chi_{xyy} & 0
\end{array}\right]$.
The nonlinear Hall angle is defined by $\theta_{non}=j_{y}/j_{x}$, which
demonstrates the propagation direction of the nonlinear AHE. By defining the
orientation of the ac electric field and the orientation of the strain $\tan\theta=E_{y}/E_{x}=\alpha/\beta$ and $\tan\phi=D_{x}/D_{y}$, the nonlinear Hall angle is straightforwardly obtained as
$\tan\theta_{non}=\left(-\cot\theta\right)n_{e}^{2}$
with
\begin{align}
n_{e}^{2} & =\frac{\tan\phi+2\tan\theta}{2\tan\phi+\tan\theta}. \label{eq8}
\end{align}
Here, $D_{x}$ and $D_{y}$ is proportional to $\alpha$ and $\beta$ respectively, and the ratio are equal. $D_{x}/D_{y}=\alpha/\beta$ due to the $C_{3v}$ symmetry.
In the former references \cite{15,18}, only the ac electric field along the $x$ direction ($y$ direction) is taken into consideration, which is $\theta = 0$ $(\theta=\pi/2)$ and the nonlinear Hall angle is $\theta_{non} = \pi/2$ $(\theta_{non} = 0)$ whatever
the strain orientation is. However, when the orientation of the ac electric field is an arbitrary
one, the nonlinear Hall angle is tuned both by the orientation of the ac electric field $\theta$ and the direction defined by the BCD moments $\phi$. Here, $n_{e}$
actually plays the same role of the refraction index in the birefraction phenomenon in optics. The cartoon of those directions according to the Huygens' Principle
is shown in Fig.~\ref{fig3}, where the radius of the circle is 1 and the major
axis and minor axis of ellipse are 1 and $\left|n_{e}\right|$ respectively. The $n_{e}^{2}>1$ (Fig.~\ref{fig3}(a)) and $0<n_{e}^{2}<1$
(Fig.~\ref{fig3}(b)) correspond to the positive and negative crystals for the
birefraction phenomenon. Besides these ordinary cases, $n_{e}^{2}<-1$
(Fig.~\ref{fig3}(c)) and $-1<n_{e}^{2}<0$ (Fig.~\ref{fig3}(d)) also exist, which have
no optical counterparts since the refraction indices become imaginary
in these two cases. The $\theta_{non}$ in imaginary positive crystal (imaginary negative crystal) is equal in magnitude, but opposite in sign with $\theta_{non}$ in positive crystal (negative crystal).

\begin{figure}[h]
\centering
\includegraphics[width=5cm]{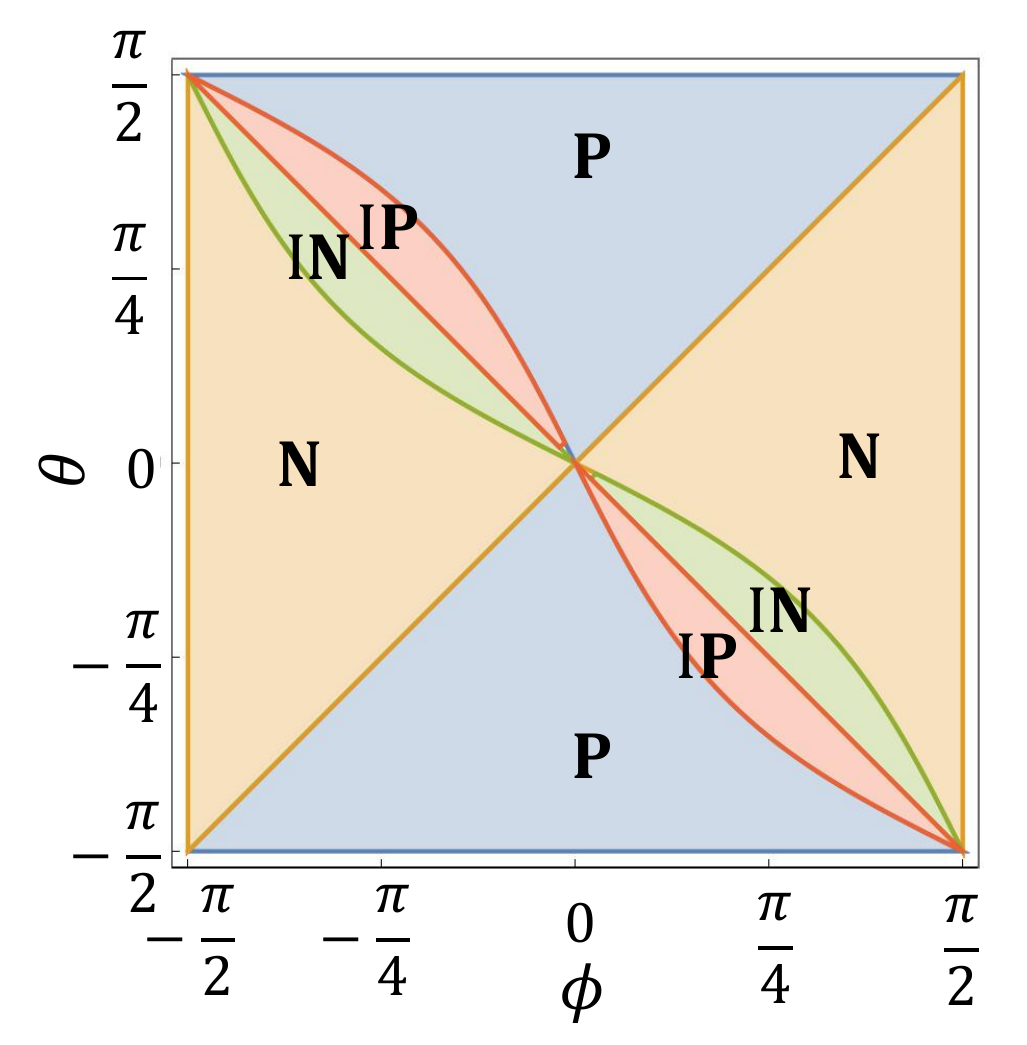}
\caption{The phase diagram of $n_{e}^{2}$ versus $\theta$
and $\phi$. The blue, yellow, red and green areas represent positive
crystal (P), negative crystal (N), the imaginary positive crystal (IP)
and imaginary negative crystal (IN) phases. }
\label{fig4}
\end{figure}

We provide the phase diagram of $n^2_{e}$ versus $\theta$ and $\phi$ in Fig.~\ref{fig4}. The
blue, yellow, red and green areas represent positive crystal (P), negative crystal (N), the imaginary positive crystal (IP) and imaginary negative crystal (IN) phases. As we can see, if $\theta$ and $\phi$ are switched, $n^2_{e}$ becomes the reciprocal of itself, and eventually the negative phases and the positive phases are correspondingly switched.
%%%%%%%%%%%%%%%%%%%%%%%%%%%%%%%%%%%%%%%%%%%%%%%%%%%%%%%%%%%%%
%%%%%%%%%%%%%%%%%%%%%%%%%%%%%%%%%%%%%%%%%%%%%%%%%%%%%%%%%%%%%

In conclusion, we study the uniaxial strain tuned nonlinear AHE
in the $\textrm{MoS}_{2}$ monolayer. By applying ac electric field and
the uniaxial strain, the nonlinear AHE can be achieved in such system.
We find that not only the magnitude of the nonlinear AHE
but also the nonlinear Hall angle can be tuned by the strain.
The nonlinear Hall angle, the orientation of the strain axis and
the orientation of the ac electric field exhibit
a deep relationship which is analogy to the birefraction phenomenon
in optics. The orientation of the ac electric field and the nonlinear Hall angle play the roles as the ordinary
and the extraordinary lights in the optical counterparts. Just like
the birefraction resulting from the dielectric constant tensor, the
hidden relationship between the orientation of the ac electric field and nonlinear Hall angles can
be established due to the pseudotensor nature of the BCD. In this sense, there are effective refraction indices for
these birefraction-like phenomenon in $\textrm{MoS}_{2}$.
Besides the ordinary positive and negative crystals in optics, we
also find that there are two more birefraction-like cases corresponding
to an imaginary refraction index in $\textrm{MoS}_{2}$  monolayer when accurately tuning the uniaxial strain. Our findings can be generalized
to other 2D systems and shed lights on the strain tuning
electronic devices based on the 2D materials.

\section{Method}
The band structure of $\textrm{MoS}_{2}$ was calculated using first-principles simulations within density functional theory (DFT). The projected augmented wave pseudopotentials method was used as implemented in the Vienna Ab $initio$ Simulation Package (VASP). The exchange correlation energy was calculated using the generalized gradient approximation (GGA) of the Perdew-Burke-Ernzerhof form, and the plane wave cutoff energy was set to 500 eV. For the calculation of band structure, a centered 21 x 21 x 1 kpoint mesh was used. The vacuum space along the z direction was set to be $>$ 20 \AA. The plane lattice constant and atomic coordinates were fully relaxed until the energy and force converged to $10^{-8}$ and $10^{-2}$ eV/\AA, respectively. 

\begin{acknowledgement}
This work was supported by NSFC Grants No. 12175150 and Guangdong Basic and Applied Basic Research Foundation (Grant No. 2022A1515012006 and 2023A1515011223.)).
\end{acknowledgement}


%% graphical table of contents.
%%%%%%%%%%%%%%%%%%%%%%%%%%%%%%%%%%%%%%%%%%%%%%%%%%%%%%%%%%%%%%%%%%%%%


\begin{thebibliography}{99}
\bibitem{1}
X. Gu, A. F. Kockum, A. Miranowicz, Y.-X. Liu, and F. Nori, ``Microwave photonics with superconducting quantum circuits,'' \textit{Phys. Rep.}, \textbf{718}, 1--102 (2017).

\bibitem{2}
D. Xiao, M.-C. Chang, and Q. Niu, ``Berry phase effects on electronic properties,'' \textit{Rev. Mod. Phys.}, \textbf{82}, 1959--2007 (2010).

\bibitem{3}
D. Xiao, W. Yao, and Q. Niu, ``Valley-contrasting physics in graphene: magnetic moment and topological transport,'' \textit{Phys. Rev. Lett.}, \textbf{99}, 236809 (2007).

\bibitem{4}
D. Xiao, G.-B. Liu, W. Feng, X. Xu, and W. Yao, ``Coupled spin and valley physics in monolayers of MoS$_2$ and other group-VI dichalcogenides,'' \textit{Phys. Rev. Lett.}, \textbf{108}, 196802 (2012).

\bibitem{5}
Y. K. Kato, R. C. Myers, A. C. Gossard, and D. D. Awschalom, ``Observation of the spin Hall effect in semiconductors,'' \textit{Science}, \textbf{306}, 1910--1913 (2004).

\bibitem{6}
J. Wunderlich, B. Kaestner, J. Sinova, and T. Jungwirth, ``Experimental observation of the spin-Hall effect in a two-dimensional spin-orbit coupled semiconductor system,'' \textit{Phys. Rev. Lett.}, \textbf{94}, 047204 (2005).

\bibitem{7}
W. Yao, D. Xiao, and Q. Niu, ``Valley-dependent optoelectronics from inversion symmetry breaking,'' \textit{Phys. Rev. B}, \textbf{77}, 235406 (2008).

\bibitem{8}
I. Souza and D. Vanderbilt, ``Dichroic f-sum rule and the orbital magnetization of crystals,'' \textit{Phys. Rev. B}, \textbf{77}, 054438 (2008).

\bibitem{9}
Z. Gong, G.-B. Liu, H. Yu, D. Xiao, X. Cui, X. Xu, and W. Yao, ``Magnetoelectric effects and valley-controlled spin quantum gates in transition metal dichalcogenide bilayers,'' \textit{Nat. Commun.}, \textbf{4}, 2053 (2013).

\bibitem{10}
G. Aivazian, Z. Gong, A. M. Jones, R.-L. Chu, J. Yan, D. G. Mandrus, C. Zhang, D. Cobden, W. Yao, and X. Xu, ``Magnetic control of valley pseudospin in monolayer WSe$_2$,'' \textit{Nat. Phys.}, \textbf{11}, 148--152 (2015).

\bibitem{11}
A. Srivastava, M. Sidler, A. V. Allain, D. S. Lembke, A. Kis, and A. Imamo\u{g}lu, ``Valley Zeeman effect in elementary optical excitations of monolayer WSe$_2$,'' \textit{Nat. Phys.}, \textbf{11}, 141--147 (2015).

\bibitem{12}
N. Nagaosa, J. Sinova, S. Onoda, A. H. MacDonald, and N. P. Ong, ``Anomalous Hall effect,'' \textit{Rev. Mod. Phys.}, \textbf{82}, 1539--1592 (2010).

\bibitem{13}
M. I. Dyakonov and A. V. Khaetskii, ``Spin Hall effect,'' in \textit{Spin Physics in Semiconductors}, pp. 211--243, Springer, 2008.

\bibitem{14}
J. Lai, J. Zhan, P. Liu, X.-Q. Chen, and Y. Sun, ``Electric field tunable nonlinear Hall terahertz detector in the dual quantum spin Hall insulator TaIrTe$_4$,'' \textit{Phys. Rev. B}, \textbf{110}, 155122 (2024).

\bibitem{15}
I. Sodemann and L. Fu, ``Quantum nonlinear Hall effect induced by Berry curvature dipole in time-reversal invariant materials,'' \textit{Phys. Rev. Lett.}, \textbf{115}, 216806 (2015).

\bibitem{16}
Z. Zhang, Z.-G. Zhu, and G. Su, ``Theory of nonlinear response for charge and spin currents,'' \textit{Phys. Rev. B}, \textbf{104}, 115140 (2021).

\bibitem{17}
A. Naseer, A. Priydarshi, P. Ghosh, R. Ahammed, Y. S. Chauhan, S. Bhowmick, and A. Agarwal, ``Room temperature ferroelectricity and an electrically tunable Berry curvature dipole in III--V monolayers,'' \textit{Nanoscale}, (2024).

\bibitem{18}
G. Kim, J. Bahng, J. Jeong, W. Sakong, T. Lee, D. Lee, Y. Kim, H. Rho, and S. C. Lim, ``Gate modulation of dissipationless nonlinear quantum geometric current in 2D Te,'' \textit{Nano Lett.}, \textbf{24}, 10820--10826 (2024).

\bibitem{19}
L. Wang, J. Zhu, H. Chen, H. Wang, J. Liu, Y.-X. Huang, B. Jiang, J. Zhao, H. Shi, G. Tian, \textit{et al.}, ``Orbital magneto-nonlinear anomalous Hall effect in kagome magnet Fe$_3$Sn$_2$,'' \textit{Phys. Rev. Lett.}, \textbf{132}, 106601 (2024).

\bibitem{20}
N. Kheirabadi and A. Langari, ``Quantum nonlinear planar Hall effect in bilayer graphene: An orbital effect of a steady in-plane magnetic field,'' \textit{Phys. Rev. B}, \textbf{106}, 245143 (2022).

\bibitem{21}
R. Habara and K. Wakabayashi, ``Nonlinear optical Hall effect of few-layered NbSe$_2$,'' \textit{Phys. Rev. Res.}, \textbf{4}, 013219 (2022).

\bibitem{22}
S. S. Samal, S. Nandy, and K. Saha, ``Nonlinear transport without spin-orbit coupling or warping in two-dimensional Dirac semimetals,'' \textit{Phys. Rev. B}, \textbf{103}, L201202 (2021).

\bibitem{23}
J. Li, H. Jin, Y. Wei, and H. Guo, ``Tunable intrinsic spin Hall conductivity in bilayer PtTe$_2$ by controlling the stacking mode,'' \textit{Phys. Rev. B}, \textbf{103}, 125403 (2021).

\bibitem{24}
H. Jin, H. Su, X. Li, Y. Yu, H. Guo, and Y. Wei, ``Strain-gated nonlinear Hall effect in two-dimensional MoSe$_2$/WSe$_2$ van der Waals heterostructure,'' \textit{Phys. Rev. B}, \textbf{104}, 195404 (2021).

\bibitem{25}
J.-S. You, S. Fang, S.-Y. Xu, E. Kaxiras, and T. Low, ``Berry curvature dipole current in the transition metal dichalcogenides family,'' \textit{Phys. Rev. B}, \textbf{98}, 121109 (2018).

\bibitem{26}
L. Du, Z. Huang, J. Zhang, F. Ye, Q. Dai, H. Deng, G. Zhang, and Z. Sun, ``Nonlinear physics of moir茅 superlattices,'' \textit{Nat. Mater.}, \textbf{23}, 1179--1192 (2024).

\bibitem{27}
R. Chen, Z. Z. Du, H.-P. Sun, H.-Z. Lu, and X. C. Xie, ``Nonlinear Hall effect on a disordered lattice,'' \textit{Phys. Rev. B}, \textbf{110}, L081301 (2024).

\bibitem{28}
M.-S. Qin, P.-F. Zhu, X.-G. Ye, W.-Z. Xu, Z.-H. Song, J. Liang, K. Liu, and Z.-M. Liao, "Strain tunable Berry curvature dipole, orbital magnetization and nonlinear Hall effect in WSe$_2$ monolayer," \textit{Chin. Phys. Lett.} \textbf{38}, 017301 (2021).

\bibitem{29}
C.-P. Zhang and K. T. Law, ``Nonlinear Hall effect in an insulator,'' \textit{Nat. Nanotechnol.}, \textbf{19}, 1432--1433 (2024).

\bibitem{30}
C.-P. Zhang, X.-J. Gao, Y.-M. Xie, H. C. Po, and K. T. Law, ``Higher-order nonlinear anomalous Hall effects induced by Berry curvature multipoles,'' \textit{Phys. Rev. B}, \textbf{107}, 115142 (2023).

\bibitem{31}
C. Ortix, ``Nonlinear Hall Effect with Time-Reversal Symmetry: Theory and Material Realizations,'' \textit{Adv. Quantum Technol.}, \textbf{4}, 2100056 (2021).

\bibitem{32}
K. Das, K. Ghorai, D. Culcer, and A. Agarwal, ``Nonlinear valley Hall effect,'' \textit{Phys. Rev. Lett.}, \textbf{132}, 096302 (2024).

\bibitem{33}
Z. Z. Du, C. M. Wang, H. P. Sun, H. Z. Lu, and X. C. Xie, ``Quantum theory of the nonlinear Hall effect,'' \textit{Nat. Commun.}, \textbf{12}, 5038 (2021).

\bibitem{34}
H. Zeng, J. Dai, W. Yao, D. Xiao, and X. Cui, "Valley polarization in MoS$_2$ monolayers by optical pumping," \textit{Nat. Nanotechnol.} \textbf{7}, 490 (2012).

\bibitem{35}
K. F. Mak, K. He, J. Shan, and T. F. Heinz, "Control of valley polarization in monolayer MoS$_2$ by optical helicity," \textit{Nat. Nanotechnol.} \textbf{7}, 494 (2012).

\bibitem{36}
A. M. Jones, H. Yu, N. J. Ghimire, S. Wu, G. Aivazian, J. S. Ross, B. Zhao, J. Yan, D. G. Mandrus, D. Xiao, \textit{et al.}, "Optical generation of excitonic valley coherence in monolayer WSe$_2$," \textit{Nat. Nanotechnol.} \textbf{8}, 634 (2013).

\bibitem{37}
T. Cao, G. Wang, W. Han, H. Ye, C. Zhu, J. Shi, Q. Niu, P. Tan, E. Wang, B. Liu, \textit{et al.}, "Valley-selective circular dichroism of monolayer molybdenum disulphide," \textit{Nat. Commun.} \textbf{3}, 887 (2012).

\bibitem{38}
Y. Araki, ``Strain-induced nonlinear spin Hall effect in topological Dirac semimetal,'' \textit{Sci. Rep.}, \textbf{8}, 15236 (2018).

\bibitem{39}
J. Wan, Y.-L. Wu, K.-Q. Chen, and X.-Q. Yu, ``Strongly enhanced nonlinear acoustic valley Hall effect in tilted Dirac materials,'' \textit{Phys. Rev. B}, \textbf{109}, L161101 (2024).

\bibitem{40}
P. A. Pantaleón, T. Low, and F. Guinea, "Tunable large Berry dipole in strained twisted bilayer graphene," \textit{Phys. Rev. B} \textbf{103}, 205403 (2021).

\bibitem{41}
D. Xiao, G.-B. Liu, W. Feng, X. Xu, and W. Yao, "Coupled spin and valley physics in monolayers of MoS$_2$ and other group-VI dichalcogenides," \textit{Phys. Rev. Lett.} \textbf{108}, 196802 (2012).

\bibitem{42}
P. A. Pantaleón, T. Low, and F. Guinea, "Tunable large Berry dipole in strained twisted bilayer graphene," \textit{Phys. Rev. B} \textbf{103}, 205403 (2021).

\bibitem{43} 
H. Wang, C. Zhang, and F. Rana, "Surface recombination limited lifetimes of photoexcited carriers in few-layer transition metal dichalcogenide MoS$_2$," \textit{Nano Lett.} \textbf{15}, 8204 (2015).

\end{thebibliography}
\end{document}